\documentclass[english,nofootinbib,nobibnotes,secnumarabic,openacc]{rstransa}
\usepackage[T1]{fontenc}
\usepackage[latin9]{luainputenc}
\usepackage{array}
\usepackage{multirow}
\usepackage{amsmath}
\usepackage{amsthm}
\usepackage{amssymb}
\usepackage{graphicx}
\usepackage{setspace}
\PassOptionsToPackage{normalem}{ulem}
\usepackage{ulem}

\makeatletter

\providecommand{\tabularnewline}{\\}

\newcommand{\lyxaddress}[1]{
\par {\raggedright #1
\vspace{1.4em}
\noindent\par}
}
  \theoremstyle{definition}
  \newtheorem{defn}{\protect\definitionname}
  \theoremstyle{plain}
  \newtheorem{prop}{\protect\propositionname}
\theoremstyle{plain}
\newtheorem{thm}{\protect\theoremname}

\makeatother

\usepackage{babel}
  \providecommand{\definitionname}{Definition}
  \providecommand{\propositionname}{Proposition}
\providecommand{\theoremname}{Theorem}

\begin{document}
\setlength{\extrarowheight}{2pt}

\title{Relating Causal and Probabilistic Approaches to Contextuality}

\author{Matt Jones$^{1}$}

\lyxaddress{\address{$^{1}$University of Colorado Boulder}}

\lyxaddress{\subject{physics, psychology, mathematics}}

\lyxaddress{\keywords{contextuality, probabilistic causal models, contextuality-by-default,
direct influence}}

\lyxaddress{\corres{Matt Jones\\
 \email{mcj@colorado.edu}}}
\begin{abstract}
\begin{singlespace}
\noindent A primary goal in recent research on contextuality has been
to extend this concept to cases of inconsistent connectedness, where
observables have different distributions in different contexts. This
article proposes a solution within the framework of probabilistic
causal models, which extend hidden-variables theories, and then demonstrates
an equivalence to the contextuality-by-default (CbD) framework. CbD
distinguishes contextuality from direct influences of context on observables,
defining the latter purely in terms of probability distributions.
Here we take a causal view of direct influences, defining direct influence
within any causal model as the probability of all latent states of
the system in which a change of context changes the outcome of a measurement.
Model-based contextuality (M-contextuality) is then defined as the
necessity of stronger direct influences to model a full system than
when considered individually. For consistently connected systems,
M-contextuality agrees with standard contextuality. For general systems,
it is proved that M-contextuality is equivalent to the property that
any model of a system must contain ``hidden influences'', meaning
direct influences that go in opposite directions for different latent
states, or equivalently signaling between observers that carries no
information. This criterion can be taken as formalizing the ``no-conspiracy''
principle that has been proposed in connection with CbD. M-contextuality
is then proved to be equivalent to CbD-contextuality, thus providing
a new interpretation of CbD-contextuality as the non-existence of
a model for a system without hidden direct influences.
\end{singlespace}
\end{abstract}
\maketitle

\section{\label{sec:Introduction}Introduction}

Probabilistic contextuality describes an empirical system of measurements
wherein a set of observables can be measured in different subsets
in different contexts, and even though each observable has the same
distribution in all contexts in which it is measured, the joint distributions
of measurements within each context cannot be pieced together into
a global joint distribution. The original significance lies in Bell's
theorem and related results \cite{bell64,CHSH}, which state that
certain contextual systems cannot be explained by any local hidden-variables
theory, although such systems are predicted by quantum mechanics and
have been experimentally confirmed \cite{aspect81,hensen15}.

The standard formulation of contextuality applies only to cases where
the distribution of each observable is identical across contexts.
This property of a system of measurements is called consistent connectedness
\cite{KuDzLa15}, marginal selectivity \cite{DzKj14,TownsendSchweickert89},
or no-disturbance \cite{Ramanathan-etal}. Dzhafarov and Kujala have
made persuasive arguments for extending contextuality to inconsistently
connected systems, including the fact that real experiments never
eliminate all sources of contamination, and that sample frequencies
in finite datasets will generally not be equal even if the true generating
probabilities are \cite{Dzh-nothing-something,DzKj15-conversations,DzKj16}.
Beyond these pragmatic considerations, it is of interest to know whether
contextuality can be usefully defined in cases where the distributions
of observables are truly different across contexts. This might enable
contextuality analysis to be applied to other domains, such as human
cognition or behavior \cite{AertsEtal18,AertsGaboraSozzo13,CervantesDz18-snowqueen,DzEtal16}.

The contextuality-by-default theory (CbD) offers one approach for
defining contextuality for inconsistently connected systems \cite{DzKj16,DzKj17-CbD2,KuDzLa15}.
CbD treats measurements of each observable in different contexts as
different random variables, by default. It then asks whether the distributions
of random variables in each context are compatible with a global distribution
in which all variables for each observable are made as equal as possible,
in a rigorous sense based on probabilistic couplings \cite{Thorisson}.
If not, then the system is CbD-contextual (we use this term to distinguish
from standard contextuality and the model-based M-contextuality introduced
below). For consistently connected systems, CbD-contextuality agrees
with standard contextuality.

In addition to applying beyond consistently connected systems, CbD
departs from previous approaches to contextuality in that it is a
purely probabilistic theory of random variables, not grounded in theories
of the physical system generating the measurements. As such, it is
unclear what CbD-contextuality indicates about that system. Does establishing
that an inconsistently connected set of measurements is CbD-contextual
imply anything about viable theories of the physical system, similar
to how standard contextuality implies a system cannot be described
by any local hidden-variable theory?

The present article proves an affirmative answer to this question,
based on a characterization of contextuality recently advanced by
Cavalcanti \cite{cavalcanti18} in terms of probabilistic causal models.
Probabilistic causal models are widely used in statistics, computer
science, machine learning, and psychology and are well suited for
situations involving stochastic latent structure \cite{jordan99,pearl00}.
They provide a useful generalization of hidden-variable theories in
physics, enabling a physical system to be described by context variables
controlled by the experimenter, unobservable variables representing
theoretical latent (hidden) states of the system, and measured observables.
Within this framework, we propose a definition of model-based contextuality
(M-contextuality) that applies to both consistently and inconsistently
connected systems, and we prove that it is equivalent both to CbD-contextuality
and to the non-existence of a certain type of model for the system
under investigation. 

Our approach builds on two of the main principles that have motivated
CbD: the distinction between contextuality and direct influence \cite{CervantesDz18-snowqueen,DzKj14,DzKj16},
and Cervantes and Dzhafarov's \emph{no-conspiracy} principle prohibiting
``hidden'' direct influences \cite{CervantesDz18-snowqueen}. Early
work on CbD showed that contextuality can be defined as a context-dependence
of the identity of random variables over and above the dependence
due to direct influences. Although the notion of direct influence
was founded on the theory of selective influences in probabilistic
causal models \cite{Dzh03}, subsequent developments of CbD have defined
direct influence in purely probabilistic terms, to refer to the difference
in an observable's distribution across different contexts \cite{CervantesDz18-snowqueen,DzKj16}.
Here we refer to such distributional differences as inconsistent connectedness,
reserving direct influence to refer to a causal effect of context
on the values of observables. Whereas inconsistent connectedness is
an empirical (statistical) property of the measurements, direct influence
as defined here is a theoretical (model-dependent) property of the
physical system. 

This article proposes a quantitative definition of direct influence
within any probabilistic causal model, as the probability of all latent
(hidden) states of the model in which a change of context changes
the value of an observable. The degree of inconsistent connectedness
of a measurement system imposes a minimal amount of direct influence
needed to model each observable in any pair of contexts in which it
is measured. We define a measurement system as M-contextual if modeling
the full system requires direct influences stronger than these minimum
values. This formalizes a proposal by Cavalcanti \cite{cavalcanti18}
that ``A causal model should not allow causal connections stronger
than needed to explain the observed deviations from the no-disturbance
condition'' (p. 6). For consistently connected systems, the minimal
direct influences are zero, and the definition of M-contextuality
coincides with that of standard contextuality.

Concerning the no-conspiracy principle, Ehtibar Dzhafarov gives the
following philosophical-level statement of the principle (personal
communication, June 2018): ``Direct influences of a reasonable substantive
theory (in physics or psychology) are not revealed in the distributional
differences only under special, precariously set circumstances. As
a rule, there are no \textquotedblleft hidden\textquotedblright{}
direct influences.'' A conceptually similar and logically weaker
principle is that of \emph{no-fine-tuning} introduced by Wood and
Spekkens \cite{WoodSpekkens15} and elaborated by Cavalcanti \cite{cavalcanti18},
which holds that empirical conditional independence between measurement
outcomes arises only when there is no causal connection: causal parameters
cannot be fine-tuned such that their effects exactly balance out.
Building on the present definition of direct influence, we formalize
the idea of hidden direct influences as direct influences that work
in opposite directions for different latent states, thus leaving the
marginal distributions of observables unaffected. We then interpret
the no-conspiracy principle as a prohibition against models with hidden
influences. The primary results of this article are proofs that M-contextuality
and CbD-contextuality, as properties of a measurement system, are
both equivalent to the non-existence of a model of that system without
hidden influences (Theorems \ref{thm:Mnoncontextual-aligned}, \ref{thm:Mcontextual-CbDcontextual},
\& \ref{thm:CbDcontextual-aligned}). 

The definition proposed here for hidden direct influence agrees with
that of non-communicating signaling given in Atmanspacher and Filk's
recent criticism of CbD \cite{AtmanspacherFilk}. Likewise, their
observation that the criterion of CbD-contextuality accounts for communicating
but not non-communicating signaling anticipates the result of the
present article that a CbD-contextual system is one that cannot be
modeled without hidden direct influences. We discuss in the concluding
section how the formalism offered here reconciles the position of
Atmanspacher and Filk with that of Dzhafarov, Kujala, and colleagues,
at least at a mathematical level. More generally, the value of the
present results is that they show a formal equivalence between three
conceptually different approaches to contextuality: (1) the assumption
underlying M-contextuality that direct influence in causal models
is limited to that implied by inconsistent connectedness, (2) the
no-conspiracy and no-fine-tuning principles, and (3) the probabilistic
couplings approach of CbD. This correspondence will hopefully facilitate
understanding and further development of both CbD and model-based
approaches to contextuality.

The remainder of this article is organized as follows. Section \ref{sec:Standard-Contextuality}
gives notation and definitions for standard contextuality. Section
\ref{sec:Causal-model-Characterization} describes causal probabilistic
models and their relationship to standard contextuality. Section \ref{sec:Direct-Influence}
defines a quantitative measure of direct influence in causal models,
defines hidden direct influences, and offers a formalization of the
no-conspiracy principle. Section \ref{sec:Model-based-Contextuality}
defines M-contextuality, proves that it agrees with standard contextuality
for consistently connected systems (Theorem \ref{thm:Mcontextual-contextual}),
and proves that regardless of consistent connectedness M-contextuality
is equivalent to the non-existence of a model without hidden influences
(Theorem \ref{thm:Mnoncontextual-aligned}). Section \ref{sec:Partitionable-systems}
recasts the preceding results for systems defined by a set of separate
observers, as in Bell scenarios, relating direct influence to signaling
among observers. Section \ref{sec:Examples} gives examples. Section
\ref{sec:Relationship-to-CbD} derives a translation between the model-based
approach and CbD and proves the final main result (Theorem \ref{thm:Mcontextual-CbDcontextual}),
that M-contextuality and CbD-contextuality are equivalent. As a corollary
(Theorem \ref{thm:CbDcontextual-aligned}), we also show that CbD-contextuality
can be given a causal interpretation, in that a CbD-contextual system
is one that is incompatible with a particular class of probabilistic
causal models, namely those without hidden influences.

\section{\label{sec:Standard-Contextuality}Standard Contextuality}
\begin{defn}[Measurement system]
A measurement system consists of a set of observables $\mathcal{Q}=\left\{ q\right\} $,
a set of possible values $\mathcal{O}_{q}$ for each observable, a
set of contexts $\mathcal{C}=\left\{ c\right\} $, a relation $\prec$
with $q\prec c$ indicating that observable $q$ is measured in context
$c$, and a set of random variables $M=\left\{ M_{q}^{c}:q\in\mathcal{Q},c\in\mathcal{C},q\prec c\right\} $.
The subset $M^{c}=\left\{ M_{q}^{c}:q\prec c\right\} $ is jointly
distributed with distribution $\mu_{c}$ for each $c$, and $M_{q}^{c}$
and $M_{q'}^{c'}$ are stochastically unrelated (i.e., are not measured
together) whenever $c\neq c'$. Note the specification of $M=\left\{ M_{q}^{c}\right\} $
determines $\mathcal{Q}$, $\left\{ \mathcal{O}_{q}\right\} $, $\mathcal{C}$,
$\prec$, and $\left\{ \mu_{c}\right\} $, and therefore we can refer
to the entire measurement system as $M$. The only technical requirements
for the present results to hold are that $\mathcal{Q}$ and $\mathcal{C}$
are both countable (i.e., no larger than the infinite set of natural
numbers) and that each $\mathcal{O}_{q}$ is Hausdorff and second-countable
(this includes finite outcome spaces, $n$-dimensional Cartesian space
$\mathbb{R}^{n}$, and separable Hilbert space).
\end{defn}
Although most literature on contextuality treats the $\mu_{c}$ as
known distributions, in empirical practice one has access only to
samples from those distributions. Therefore one might argue we should
refer not to random variables $M_{q}^{c}$ but to individual observations,
$M_{q}^{c,i}$, where $i$ ($1\leq i\leq n_{c}$) indexes the instances
in which the experiment was performed in condition $c$. One advantage
of the model-based approach is that it explicitly distinguishes the
physical measurements $M$ from theoretical random variables (denoted
$F_{q}$ below) used to model those measurements. This distinction
makes the model-based approach naturally suited to handling sampling
error, by standard model-evaluation methods of null-hypothesis significance
testing or Bayesian model comparison. For ease of exposition, we set
aside sampling error for the majority of the article, treating the
$\mu_{c}$ as exactly known and referring to random variables $M_{q}^{c}$
rather than specific measurements $M_{q}^{c,i}$, and comment on model
fitting and evaluation in the concluding section.
\begin{defn}[Consistent connectedness]
A measurement system $M$ is consistently connected if each observable
has the same marginal distribution within every context in which it
is measured. That is, $M_{q}^{c}\sim M_{q}^{c'}$ whenever $q\prec c,c'$,
where $\sim$ indicates agreement in distribution.
\end{defn}
\begin{defn}[Standard contextuality]
\label{def:standard-contextuality}A measurement system is contextual
in the standard sense if it is consistently connected but the distributions
$\mu_{c}$ are not compatible with a joint distribution over all the
observables. More precisely, each $\mu_{c}$ is a probability measure
on the Cartesian product $\prod_{q\prec c}\mathcal{O}_{q}$. A joint
distribution $\mu$ over all the observables is a probability measure
on $\prod_{q\in\mathcal{Q}}\mathcal{O}_{q}$, and for each $c$ it
implies a marginal distribution on the observables measured in that
context, given by the push-forward measure $\pi_{*}^{c}\left(\mu\right)$
where $\pi^{c}$ is the natural projection $\prod_{q\in\mathcal{Q}}\mathcal{O}_{q}\rightarrow\prod_{q\prec c}\mathcal{O}_{q}$.
If there exists a $\mu$ such that $\pi_{*}^{c}\left(\mu\right)=\mu_{c}$
for all $c$, then the system is noncontextual; otherwise it is contextual.
\end{defn}
The intuitive idea behind contextuality is that the distribution of
each observable is unaffected by the context (consistent connectedness),
but nevertheless context exerts some sort of effect on the observables
that prevents them from being pieced together into a single jointly
distributed system. Importantly, the existence of a global distribution
in the sense of Definition \ref{def:standard-contextuality} immediately
implies the system is consistently connected. Therefore consistent
connectedness is a necessary property of any traditionally noncontextual
system. For a system that is inconsistently connected, the standard
notion of contextuality does not apply.

\section{\label{sec:Causal-model-Characterization}Causal-model Characterization
of Contextuality}

Following Cavalcanti \cite{cavalcanti18}, we analyze contextuality
of a measurement system in terms of how it can be explained by probabilistic
causal models \cite{pearl00}.
\begin{defn}[Causal probabilistic model]
A causal probabilistic model is a set of jointly distributed random
variables $\mathcal{X}=\left\{ X_{i}\right\} $, with a dependency
structure whereby each variable $X_{i}$ has a set of parents denoted
$Pa\left(X_{i}\right)\subset\mathcal{X}$ (possibly $Pa\left(X_{i}\right)=\emptyset$).
The relation between $X_{i}$ and $X_{i'}$ given by $X_{i}\in Pa\left(X_{i'}\right)$
defines a directed acyclic graph, and the model's joint distribution
factors as $\Pr\left[\mathcal{X}\right]=\prod_{i}\Pr\left[X_{i}\middle\vert Pa\left(X_{i}\right)\right]$. 
\end{defn}
For present purposes, we are interested in models of the physical
system that generates some set of measurements $M$. For any such
model, we can classify its variables into three types: variables the
experimenter sets (context), variables that are measured (observables),
and unobserved latent variables representing theoretical constructs
of the model. This classification enables a causal model of a physical
system to be put into a simple canonical form that underlies most
of the analysis in this article, as follows. Context variables can
be collected into a single variable $C$ ranging over the set of contexts
$\mathcal{C}$, with $Pa\left(C\right)=\emptyset$ because context
is an independent variable set by the experimenter. Observable variables
are denoted $F_{q}$ for each observable $q$. Those latent variables
that do not depend on anything else in the model, $\left\{ X_{i}:Pa\left(X_{i}\right)=\emptyset\right\} \setminus\left\{ C\right\} $,
can be collected into a single random variable $\Lambda$, which we
variously refer to as the source state, latent state, or hidden state
of the system prior to measurement. Any other latent variables, meaning
intermediate ones that depend on $\Lambda$ and mediating ones that
depend on $C$ or $\left\{ F_{q}\right\} $, may play an explanatory
role in interpreting the theory but are unnecessary for predictions,
$\Pr\left[\left\{ F_{q}\right\} \middle\vert C\right]$. That is,
we can marginalize over these other variables, so that $\Lambda$
encompasses all the internal mechanisms one might theorize for the
physical system, on which the observables might depend. The result
is a model described fully by $\Lambda$, $C$, and $\left\{ F_{q}\right\} $. 

\begin{figure}
\begin{centering}
\includegraphics[width=1\textwidth]{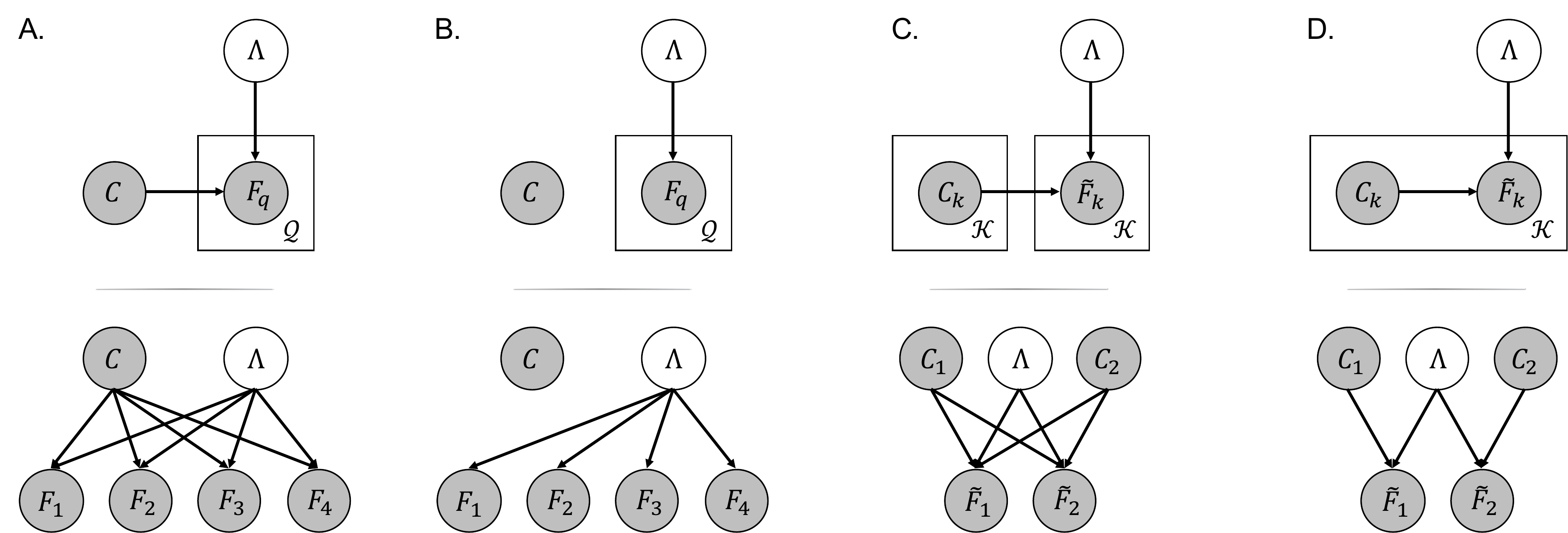}
\par\end{centering}
\caption{\label{fig:model-structure}A: Structure of a general canonical causal
model. B: Structure of a context-free canonical model. C: Structure
of a general partitioned model. D: Structure of a partitioned model
with no signaling. Filled circles represent observed variables, unfilled
circles represent unobserved variables, and arrows represent dependencies.
Upper diagram in each case uses plate notation: The variable in a
plate is replicated over the indexing set in the lower right (e.g.,
a different $F_{q}$ for every $q\in\mathcal{Q}$), and an arrow into
or out of a plate represents an arrow to or from every copy. Lower
diagram in each case explicitly depicts a model for a system with
four observables or two observers, such as the examples in Section
\ref{sec:Examples}.}
\end{figure}

This canonical structure can be further simplified in two ways. First,
because the determination of context is assumed to be under control
of the experimenter and only distributions conditioned on $C$ are
of interest, we can dispense with $C$ as a random variable and treat
it more simply as an index variable, meaning without any probabilities
associated to its values. Second, we assume each $F_{q}$ is a deterministic
function of $\Lambda$ and $C$. This assumption incurs no loss of
expressive power because one can always incorporate all stochasticity
in the model into the definition of $\Lambda$, by replacing it with
the underlying sample space. More precisely, because all variables
in a probabilistic model are jointly distributed, they can be described
as functions on a probability space $\left(\Omega,\Sigma,P\right)$.
Then $\Lambda$ is a function on $\Omega$ and each $F_{q}$ is a
function on $\Omega\times\mathcal{C}$. By replacing $\Lambda$ with
the identity function on $\Omega$, we can write each $F_{q}$ as
a function of $\Lambda$ and $C$. We call models of the resulting
structure canonical causal models and use them as our primary focus
here.
\begin{defn}[Canonical causal model]
A canonical causal model (or simply canonical model) comprises a
random variable $\Lambda$ representing the hidden state of the system
being modeled, an index variable $C$ representing the contexts in
which measurements can be made, and a set of functions $F_{q}\left(\Lambda,C\right)$
taking values in $\mathcal{O}_{q}$ and representing the measurement
outcome for each observable. The dependency structure is thus $Pa\left(F_{q}\right)=\left\{ \Lambda,C\right\} $,
$Pa\left(\Lambda\right)=Pa\left(C\right)=\emptyset$, as shown in
Figure \ref{fig:model-structure}a.
\end{defn}
It should be apparent that canonical causal models extend the class
of hidden-variables models used in classic work on contextuality (e.g.,
\cite{bell64}), by allowing context to directly influence the observables
(i.e., $C\in Pa\left(F_{q}\right)$). Prohibiting such dependencies
yields the class of (noncontextual) hidden-variables models, which
we refer to here as context-free models.
\begin{defn}[Context-free causal model]
A context-free causal model (or simply context-free model) is a canonical
model in which each $F_{q}$ is independent of $C$. That is, $Pa\left(F_{q}\right)=\left\{ \Lambda\right\} $,
and $F_{q}\left(\lambda,c\right)=F_{q}\left(\lambda\right)$ for all
$c$. See Figure \ref{fig:model-structure}b.
\end{defn}

We can now formalize the relationship between canonical models and
measurement systems:
\begin{defn}[Model for a system]
A canonical causal model $\mathcal{M}$ is a model for a measurement
system $M$ if it matches all of the individual contexts' data distributions.
That is, for each context $c$, $\Pr\left[\left\{ F_{q}:q\prec c\right\} \middle\vert C=c\right]=\mu_{c}$,
where the equality here is an equality of distributions, as probability
measures on $\prod_{q\prec c}\mathcal{O}_{q}$.
\end{defn}
The definition of a model for a system is closely related to the concept
of a coupling as used in CbD \cite{KuDzLa15}. A coupling for $M$
is a set of jointly distributed random variables $T=\left\{ T_{q}^{c}:q\in\mathcal{Q},c\in\mathcal{C},q\prec c\right\} $
such that the subset $T^{c}=\left\{ T_{q}^{c}:q\prec c\right\} $
is distributed according to $\mu_{c}$ for each $c$ (see Definition
\ref{def:probabilistic-coupling} in Section \ref{sec:Relationship-to-CbD}).
We prove in Section \ref{sec:Relationship-to-CbD} (Proposition \ref{prop:coupling-model})
that there exists a natural translation between models and couplings
for any measurement system that preserves their essential properties
regarding contextuality. However, we suggest the model-based approach
offers two conceptual advantages. First, the model-based approach
emphasizes the ontological distinction between the measurements $M_{q}^{c}$
as physical events, and the random variables $F_{q}$ as theoretical
constructs meant to explain those physical events (when taken together
with the other components of $\mathcal{M}$) \cite{cavalcanti18}.
Second, a canonical causal model contains explicit causal structure
via the latent variable $\Lambda$, which enables formal definition
of direct influences of $C$ on $F_{q}$ (see Section \ref{sec:Direct-Influence}).
This in turn enables us to distinguish direct influence, as a theoretical
property of the physical system and its dynamics (i.e., of the process
generating the data), from inconsistent connectedness, as a purely
statistical property of the data distributions.

The following three results summarize the relationship between causal
models and standard contextuality. Proposition \ref{prop:universality-measurement-models}
states that the framework of canonical causal models constitutes a
universal language capable of describing any measurement system (contextual
or not). Proposition \ref{prop:noncontextual-contextfree} recapitulates
Fine's theorem \cite{fine82} that noncontextuality is equivalent
to compatibility with a hidden-variable theory (i.e., a context-free
model). Proposition \ref{prop:CC-contextfree} states that consistent
connectedness is equivalent to the analogous property of the individual
observables. The proofs of these and all subsequent propositions and
theorems are provided in the Supplementary Material.
\begin{prop}
\label{prop:universality-measurement-models}For any measurement system
$M$, there exists a canonical causal model $\mathcal{M}$ such that
$\mathcal{M}$ is a model for $M$.
\end{prop}
\begin{prop}[Fine \cite{fine82}]
\label{prop:noncontextual-contextfree} A measurement system is noncontextual
iff there exists a context-free model of that system.
\end{prop}
\begin{prop}
\label{prop:CC-contextfree}A measurement system $M$ is consistently
connected iff there exists a context-free model for the single-observable
subsystem $M_{q}=\left\{ M_{q}^{c}:q\prec c\right\} $ for all $q$.
\end{prop}

\section{\label{sec:Direct-Influence}Direct Influence in Causal Models}

Proposition \ref{prop:noncontextual-contextfree} shows how standard
contextuality can be understood in the framework of causal models.
If a system of measurements can be explained by a causal model in
which the value of each observable depends only on the state of the
physical system prior to measurement, and not on the context (including
which other observables are measured), then the system is noncontextual.
If a measurement system is consistently connected but cannot be modeled
without assuming the observables depend on context, then it is contextual.
Furthermore, Proposition \ref{prop:CC-contextfree} shows that consistent
connectedness has a similar relationship to models of individual observables:
A system is consistently connected iff each separate observable can
be modeled in a way that it is not dependent on context. Therefore
a contextual system is one in which some dependence of the observables
on the context is required to model the full system, even though no
such dependence is needed to model any observable on its own. 

These observations suggest an extension of the concept of contextuality
to inconsistently connected systems. Specifically, we define a formal
measure of \emph{direct influence} of $C$ on each $F_{q}$ within
any canonical causal model. We then propose to define any measurement
system as contextual if modeling the complete system requires one
or more direct influences to be greater than is necessary when they
are considered separately. Zero direct influence will correspond to
a context-free model. Therefore our extended definition of contextuality
coincides with the standard one for the case of consistently connected
systems: In that case, each separate observable can be modeled with
zero direct influence, and the system is contextual iff the full system
can also be modeled with zero direct influence.

In a context-free model, a change between two contexts never changes
the outcome, regardless of the state of the physical system: $F_{q}\left(\lambda,c\right)=F_{q}\left(\lambda,c'\right)$
for all $q$, $\lambda$, and $c,c'\succ q$. We build on this property
to define a quantitative measure of direct influence in any canonical
causal model.
\begin{defn}[Direct influence]
 Given a canonical model $\mathcal{M}=\left(\Lambda,C,\left\{ F_{q}:q\in\mathcal{Q}\right\} \right)$,
the direct influence on each $F_{q}$ for any pair of contexts $c$
and $c'$ is defined as $\Delta_{c,c'}\left(F_{q}\right)=\Pr\left[\left\{ \lambda:F_{q}\left(\lambda,c\right)\neq F_{q}\left(\lambda,c'\right)\right\} \right]$.
Thus $\Delta_{c,c'}\left(F_{q}\right)$ represents the probability
that a change of context between $c$ and $c'$ would change the value
of observable $q$, with respect to the distribution over hidden states.
\end{defn}
It is important to note that direct influence as defined here is a
characteristic of a model, not of the set of measurements being modeled.
Thus the direct influence attributed to a system is a theoretically
relative construct (i.e., it depends on one's theory of the underlying
physical system), as opposed to consistent connectedness which is
a purely empirical property of the measurements alone. Nevertheless,
we can use direct influence to define contextuality, a property of
the measurement system alone, by quantifying over models (see Section
\ref{sec:Model-based-Contextuality}).

An important property related to direct influence is whether a model
contains direct influences in opposing directions, which we refer
to as hidden influences. We provide here the definition as it applies
to observables with countably many possible values. The general definition,
provided in the Supplementary Material, encompasses the definition
here and is conceptually similar but more technical.
\begin{defn}[Aligned model vs. hidden influences]
\label{def:hidden-influence}A canonical model $\mathcal{M}=\left(\Lambda,C,\left\{ F_{q}:q\in\mathcal{Q}\right\} \right)$
is aligned if, for any observable $q$, value $v\in\mathcal{O}_{q}$,
and pair of contexts $c,c'\succ q$, either $\Pr\left[\left\{ \lambda:F_{q}\left(\lambda,c\right)=v,F_{q}\left(\lambda,c'\right)\neq v\right\} \right]=0$
or $\Pr\left[\left\{ \lambda:F_{q}\left(\lambda,c\right)\neq v,F_{q}\left(\lambda,c'\right)=v\right\} \right]=0$.
If, alternatively, both of these sets have positive probability (for
some $q$, $v$, $c$, $c'$), we say the model contains hidden influences.
That is, the switch from context $c$ to $c'$ changes the value of
observable $q$ to $v$ for some states and away from $v$ for other
states.
\end{defn}
Given this definition, Cervantes and Dzhafarov's no-conspiracy principle
\cite{CervantesDz18-snowqueen} can be formalized as a prohibition
against models with hidden direct influences, or equivalently a restriction
to aligned models.

\section{\label{sec:Model-based-Contextuality}Model-based Contextuality}

Given the formal measure defined above of direct influence of context
upon an observable within a canonical causal model, we can now propose
a definition of model-based contextuality that applies to inconsistently
connected systems as well as to consistently connected ones. If a
system is inconsistently connected, one can consider how great each
direct influence must be, to model the difference in distribution
for any observable between any pair of contexts. One can then ask
whether these minimal direct influences are mutually compatible: That
is, does modeling the full system require direct influences to be
greater than is necessary individually? This approach formalizes the
generalized no-fine-tuning principle proposed by Cavalcanti \cite{cavalcanti18},
that direct influences should be no greater than required by violations
of consistent connectedness.
\begin{defn}[M-contextuality]
A measurement system $M$ is M-noncontextual if there exists a canonical
model $\mathcal{M}$ for $M$ that simultaneously minimizes all direct
influences. That is, for each $q$ and $c,c'\succ q$, $\mathcal{M}$
achieves the minimum value of $\Delta_{c,c'}\left(F_{q}\right)$ over
all models for $M$. If such a model does not exist, $M$ is M-contextual.
\end{defn}
For consistently connected systems, this definition reduces to that
of standard contextuality. Indeed, the minimal direct influences for
a consistently connected system are all zero, and consequently M-contextuality
becomes equivalent to the nonexistence of a context-free model, which
is equivalent to standard contextuality by Proposition \ref{prop:noncontextual-contextfree}.
\begin{thm}
\label{thm:Mcontextual-contextual}For consistently connected systems,
M-contextuality is equivalent to standard contextuality.
\end{thm}
Our main result regarding M-contextuality is a characterization in
terms of the existence of aligned models. The proof is based on explicit
determination of the lower bound on direct influence for each observable
and context pair and on showing that the models meeting that bound
are precisely those without hidden direct influences.
\begin{thm}
\label{thm:Mnoncontextual-aligned}A measurement system $M$ is M-contextual
iff there does not exist an aligned canonical model for $M$.
\end{thm}
According to Theorem \ref{thm:Mnoncontextual-aligned}, an M-contextual
system is one for which all models must contain hidden direct influences.
Thus M-contextuality embodies the no-conspiracy principle: If one
excludes hidden influences \emph{a priori}, then the criterion of
M-noncontextuality is simply that a system can be modeled by a classical
probabilistic model. Put differently, if there is no aligned model
for a system, then, following the logic of the no-conspiracy principle
\cite{CervantesDz18-snowqueen}, we conclude the system cannot be
explained by direct influences alone and that therefore there are
contextual influences in addition to the direct ones.

\section{\label{sec:Partitionable-systems}Partitionable systems}

In many measurement systems of interest, the observables can be partitioned
such that exactly one observable from each subset is measured in any
context. This situation arises in quantum physics when there are multiple
observers and each observer can measure one out of a set of pairwise
incompatible observables, as in Bell scenarios.
\begin{defn}[Partitionable measurement system]
A partitionable measurement system is one in which the set of observables
can be partitioned as $\mathcal{Q}=\bigsqcup_{k\in\mathcal{K}}\mathcal{Q}_{k}$,
the set of contexts is $\mathcal{C}=\prod_{k\in\mathcal{K}}\mathcal{Q}_{k}$,
and $q\prec c$ iff $c_{k}=q$ for each $q\in\mathcal{Q}_{k}$. That
is, every context $c$ corresponds to a choice of exactly one observable
(denoted $c_{k}$) from each subset $\mathcal{Q}_{k}$. We say $k$
indexes observers, and $\mathcal{Q}_{k}$ is the set of (pairwise
incompatible) observables available to observer $k$.
\end{defn}
When a measurement system is partitionable, it admits an alternative
form of causal model, with one outcome variable per observer that
depends on which measurement that observer chooses as well as (potentially)
on the choices of all other observers. We refer to this as a partitioned
model.
\begin{defn}[Partitioned model]
\label{def:partitioned-model}Given a partitionable measurement system
$M$ and a canonical model $\mathcal{M}=\left(\Lambda,C,\left\{ F_{q}\right\} \right)$
for $M$, the corresponding partitioned model $\tilde{\mathcal{M}}=\left(\Lambda,\left\{ C_{k}\right\} ,\left\{ \tilde{F}_{k}\right\} \right)$
is defined as follows. The hidden state of the system is modeled by
the same random variable, $\Lambda$, as in $\mathcal{M}$. The context
is decomposed into a set of variables $\left\{ C_{k}:k\in\mathcal{K}\right\} $,
with $C_{k}$ ranging over $\mathcal{Q}_{k}$ and indicating which
observable is measured by observer $k$. The measurement outcome for
observer $k$ is represented by a variable $\tilde{F}_{k}$ defined
by $\tilde{F}_{k}\left(\lambda,c\right)=F_{c_{k}}\left(\lambda,c\right)$. 
\end{defn}
The general dependency structure of $\tilde{\mathcal{M}}$ is $Pa\left(\Lambda\right)=Pa\left(C_{k}\right)=\emptyset$
and $Pa\left(\tilde{F}_{k}\right)=\left\{ \Lambda\right\} \cup\left\{ C_{k'}:k'\in\mathcal{K}\right\} $,
as illustrated in Figure \ref{fig:model-structure}c. Note that the
range of $\tilde{F}_{k}$ is the union of outcome spaces $\bigcup_{q\in\mathcal{Q}_{k}}\mathcal{O}_{q}$,
although the image of $\tilde{F}_{k}\left(\cdot,c\right)$ is contained
in $\mathcal{O}_{c_{k}}$ for any $c$. Therefore $\Pr\left[\left\{ \tilde{F}_{k}\right\} \middle\vert C=c\right]$
can be taken as a distribution on $\prod_{k}\mathcal{O}_{c_{k}}$,
which under Definition \ref{def:partitioned-model} matches the distribution
$\Pr\left[\left\{ F_{q}:q\prec c\right\} \middle\vert C=c\right]$
from $\mathcal{M}$. Therefore $\tilde{\mathcal{M}}$ is a model for
$M$ whenever $\mathcal{M}$ is. 

A partitioned model $\tilde{\mathcal{M}}$ expresses the same causal
theory of a system as the corresponding canonical model $\mathcal{M}$,
and therefore notions of direct influence and aligned models can be
straightforwardly translated to partitioned models. The direct influence
of interest here is the influence on one observer's outcome due to
the other observers' choices of observables, which we refer to as
signaling.
\begin{defn}[Signaling]
Given a partitioned model $\tilde{\mathcal{M}}$, an observer $k$,
and contexts $c$ and $c'$ with $c_{k}=c'_{k}$, signaling to observer
$k$ is defined as $\tilde{\Delta}_{c,c'}\left(\tilde{F}_{k}\right)=\Pr\left[\left\{ \lambda:\tilde{F}_{k}\left(\lambda,c\right)\neq\tilde{F}_{k}\left(\lambda,c'\right)\right\} \right]$.
That is, $\tilde{\Delta}_{c,c'}\left(\tilde{F}_{k}\right)$ is the
probability of a latent state in which switching the measurement choices
of other observers (i.e., changing $C_{k'}$ for one or more $k'\neq k$)
changes the outcome for observer $k$.
\end{defn}
It is easy to see that signaling in a partitioned model $\tilde{\mathcal{M}}$
is equal to direct influence in the corresponding canonical model
$\mathcal{M}$. That is, $\tilde{\Delta}_{c,c'}\left(\tilde{F}_{k}\right)=\Delta_{c,c'}\left(F_{q}\right)$
whenever $c_{k}=c'_{k}=q$. We can also define hidden signals versus
aligned partitioned models, paralleling the definition of hidden direct
influences and aligned canonical models. As with Definition \ref{def:hidden-influence},
the definition given here applies to observables with countably many
possible values; the general definition is provided in the Supplementary
Material.
\begin{defn}[Aligned vs. hidden signals]
\label{def:hidden-signals}A partitioned model $\tilde{\mathcal{M}}$
is aligned if, for any observer $k$, value $v\in\bigcup_{q\in\mathcal{Q}_{k}}\mathcal{O}_{q}$,
and pair of contexts $c$ and $c'$ with $c_{k}=c'_{k}$, either $\Pr\left[\left\{ \lambda:\tilde{F}_{k}\left(\lambda,c\right)=v,\tilde{F}_{k}\left(\lambda,c'\right)\neq v\right\} \right]=0$
or $\Pr\left[\left\{ \lambda:\tilde{F}_{k}\left(\lambda,c\right)\neq v,\tilde{F}_{k}\left(\lambda,c'\right)=v\right\} \right]=0$.
If, alternatively, both sets have positive probability (for some $k$,
$v$, $c$, $c'$), we say $\tilde{\mathcal{M}}$ contains hidden
signals. That is, a specific change of choices of observables for
other observers can change observer $k$'s measurement outcome to
$v$ for some states and away from $v$ for other states.
\end{defn}
The distinction between aligned and hidden signals is the same as
that between communicating and non-communicating signals \cite{AtmanspacherFilk},
because the aligned signals affect the marginal distribution of the
receiving observer's ($k$'s) measurements, whereas purely hidden
signals do not. It is easy to see that a canonical model $\mathcal{M}$
is aligned iff the corresponding partitioned model $\tilde{\mathcal{M}}$
is aligned. Likewise, $\mathcal{M}$ is context-free iff $\tilde{\mathcal{M}}$
has no signaling. The latter condition means that $\tilde{F}_{k}\left(\lambda,c\right)=\tilde{F}_{k}\left(\lambda,c'\right)$
whenever $c_{k}=c'_{k}$, so that each observer's outcome can be written
as $\tilde{F}_{k}\left(\lambda,c_{k}\right)$, and $\tilde{\mathcal{M}}$
conforms to the simplified dependency structure $Pa\left(\tilde{F}_{k}\right)=\left\{ \Lambda,C_{k}\right\} $
for every $k$ (see Figure \ref{fig:model-structure}d). These observations
imply a partitioned analogue of Proposition \ref{prop:noncontextual-contextfree},
whereby standard contextuality for consistently connected partitionable
systems is equivalent to the nonexistence of a partitioned model without
signaling, as in the original analyses of Bell scenarios \cite{CHSH}:
\begin{prop}
\label{prop:noncontextual-nosignaling}A consistently connected partitionable
measurement system $M$ is noncontextual iff there exists a partitioned
model for $M$ that has no signaling.
\end{prop}
The definition of M-contextuality can also be expressed in terms of
partitioned models:
\begin{thm}
\label{thm:partitioned-Mcontextuality}A partitionable measurement
system $M$ is M-noncontextual iff there exists a partitioned model
$\tilde{\mathcal{M}}$ that simultaneously minimizes all signaling.
That is, for each $k$, $c$, and $c'$ with $c_{k}=c'_{k}$, $\tilde{\mathcal{M}}$
achieves the minimum value of $\tilde{\Delta}_{c,c'}\left(\tilde{F}_{k}\right)$
over all partitioned models for $M$.
\end{thm}
Finally, the analogue of Theorem \ref{thm:Mnoncontextual-aligned}
for partitionable systems states that M-contextuality is equivalent
to the nonexistence of a model without hidden signaling:
\begin{thm}
\label{thm:partitionable-aligned}A partitionable measurement system
$M$ is M-contextual iff there does not exist an aligned partitioned
model for $M$.
\end{thm}

\section{\label{sec:Examples}Examples}

\paragraph*{PR box}

The Popescu-Rohrlich (PR) box \cite{Popescu-Rohrlich} is a measurement
system with four binary observables and four contexts, $\left\{ M_{1}^{1},M_{2}^{1},M_{2}^{2},M_{3}^{2},M_{3}^{3},M_{4}^{3},M_{4}^{4},M_{1}^{4}\right\} $,
with marginals $\Pr\left[M_{q}^{c}=1\right]=\frac{1}{2}$ for all
$q\prec c$ (Table \ref{tab:PR-box-marginals}), and correlations
$\Pr\left[M_{1}^{1}=M_{2}^{1}\right]=1$, $\Pr\left[M_{2}^{2}=M_{3}^{2}\right]=1$,
$\Pr\left[M_{3}^{3}=M_{4}^{3}\right]=1$, $\Pr\left[M_{1}^{4}=M_{4}^{4}\right]=-1$
(Table \ref{tab:PR-box-correlations}). This system is consistently
connected and contextual (hence also M-contextual). Consequently,
the measurements for each observable, $M_{q}$, can be modeled without
direct influence, but the system as a whole cannot.

\begin{table}
\begin{centering}
\begin{tabular}{ccccc}
\hline 
$\Pr\left[M_{q}^{c}=1\right]$ & \multicolumn{4}{c}{$q$}\tabularnewline
\cline{2-5} 
$c$ & 1 & 2 & 3 & 4\tabularnewline
\hline 
\multirow{1}{*}{$1$} & $\tfrac{1}{2}$ & $\tfrac{1}{2}$ &  & \tabularnewline
$2$ &  & $\tfrac{1}{2}$ & $\tfrac{1}{2}$ & \tabularnewline
$3$ &  &  & $\tfrac{1}{2}$ & $\tfrac{1}{2}$\tabularnewline
$4$ & $\tfrac{1}{2}$ &  &  & $\tfrac{1}{2}$\tabularnewline
\hline 
\end{tabular}
\par\end{centering}
\caption{\label{tab:PR-box-marginals}Marginal distributions for the PR box.
Each observable takes values in $\left\{ -1,1\right\} $.}
\end{table}
\begin{table}
\begin{centering}
\begin{tabular}{ccc}
\hline 
$\Pr\left[M_{q}^{c}=M_{q'}^{c}\right]$ & \multicolumn{2}{c}{$q'$}\tabularnewline
\cline{2-3} 
$q$ & $2$ & $4$\tabularnewline
\hline 
$1$ & $1$ & $-1$\tabularnewline
$3$ & $1$ & $1$\tabularnewline
\hline 
\end{tabular}
\par\end{centering}
\caption{\label{tab:PR-box-correlations}Correlation structure of the PR box.
For each pair of observables ($q,q'$) that can be measured together,
the table shows their probability of equality within the corresponding
context $(c$).}
\end{table}

Define a model $\mathcal{M}$ for the PR box by taking $\Lambda$
to range over $\left\{ -1,1\right\} $ with uniform distribution,
and defining the $F_{q}$ by
\[
\begin{array}{cccc}
F_{1}\left(\lambda,1\right)=\lambda & F_{2}\left(\lambda,2\right)=\lambda & F_{3}\left(\lambda,3\right)=\lambda & F_{4}\left(\lambda,4\right)=-\lambda\\
F_{2}\left(\lambda,1\right)=\lambda & F_{3}\left(\lambda,2\right)=\lambda & F_{4}\left(\lambda,3\right)=\lambda & F_{1}\left(\lambda,4\right)=\lambda.
\end{array}
\]
Then $\Delta_{c,c'}\left(F_{q}\right)=0$ for $q=1,2,3$, but $\Delta_{c,c'}\left(F_{4}\right)=1$
(where $c$ and $c'$ are the two applicable contexts for each $q$).
Furthermore, the direct influence on $F_{4}$ is hidden: For $\Lambda=1$,
a switch from context $3$ to context $4$ changes $F_{4}$ from $1$
to $-1$, whereas for $\Lambda=-1$ the same switch of context changes
$F_{4}$ from $-1$ to $1$. These opposing influences cancel out
in the margin. This situation illustrates the general fact that causal
models for contextual systems always exist, but they must include
hidden direct influences.

The PR box is a partitionable system, and can be written as $\mathcal{K}=\left\{ 1,2\right\} $
and $\mathcal{Q}_{1}=\left\{ 1,3\right\} $, $\mathcal{Q}_{2}=\left\{ 2,4\right\} $.
That is, observer $1$ measures observable $1$ or $3$, and observer
$2$ measures observable $2$ or $4$. The contexts are now denoted
$\mathcal{C}=\left\{ \left(1,2\right),\left(3,2\right),\left(3,4\right),\left(1,4\right)\right\} $.
The model $\mathcal{M}$ corresponds to a partitioned model $\tilde{\mathcal{M}}$
with outcome variables defined by
\[
\begin{array}{cccc}
\tilde{F}_{1}\left(\lambda,\left(1,2\right)\right)=\lambda & \tilde{F}_{1}\left(\lambda,\left(3,2\right)\right)=\lambda & \tilde{F}_{1}\left(\lambda,\left(3,4\right)\right)=\lambda & \tilde{F}_{1}\left(\lambda,\left(1,4\right)\right)=\lambda\\
\tilde{F}_{2}\left(\lambda,\left(1,2\right)\right)=\lambda & \tilde{F}_{2}\left(\lambda,\left(3,2\right)\right)=\lambda & \tilde{F}_{2}\left(\lambda,\left(3,4\right)\right)=\lambda & \tilde{F}_{2}\left(\lambda,\left(1,4\right)\right)=-\lambda.
\end{array}
\]
Signaling in this model is zero everywhere except for $\tilde{\Delta}_{\left(1,4\right),\left(3,4\right)}\left(\tilde{F}_{2}\right)=1$.
That is, observer $1$'s measurement outcome is unaffected by observer
$2$'s choice of observable, regardless of the state of the physical
system, and a similar statement holds for observer 2 when she measures
observable 2. However, when observer $2$ measures observable $4$,
the outcome depends on which variable observer $1$ measures. Moreover,
this signaling goes in opposite directions depending on the latent
state $\Lambda$. That is, $\tilde{\mathcal{M}}$ contains signaling,
and that signaling is (perfectly) hidden. In the language of Atmanspacher
and Filk \cite{AtmanspacherFilk}, this is noncommunicating signaling
that cannot transmit information between observers.

\paragraph*{An M-noncontextual inconsistently connected system}

Consider a system with the same observables, contexts, and correlation
structure as the PR box, but with unbalanced marginals as shown in
Table \ref{tab:Example-2-Marginals}. The inconsistent connectedness
of this system places a lower bound on the direct influences in any
model thereof, which can be shown to be $\Delta_{1,4}\left(F_{1}\right)\geq\frac{1}{6}$,
$\Delta_{1,2}\left(F_{2}\right)\geq\frac{1}{3}$, $\Delta_{2,3}\left(F_{3}\right)\geq\frac{1}{3}$,
and $\Delta_{3,4}\left(F_{4}\right)\geq\frac{1}{6}$. The question
is whether these lower bounds can be achieved simultaneously, and
the answer is affirmative. Let $\mathcal{M}$ be the model for this
system defined by $\Lambda$ ranging over $\left\{ 1,2,3,4,5,6\right\} $
with uniform distribution, and the $F_{q}$ as given in Table \ref{tab:Example-2-model}.
This model is immediately seen to achieve the lower bound on all $\Delta_{c,c'}\left(F_{q}\right)$.
Note also that the direct influences are all aligned. For example,
switching from context $1$ to context $2$ can change $F_{2}$ from
$1$ to $-1$ (when $\Lambda\in\left\{ 3,4\right\} $), but it cannot
change $F_{2}$ from $-1$ to $1$. These properties of $\mathcal{M}$
show the system is M-noncontextual.

The partitioned model $\tilde{\mathcal{M}}$ corresponding to $\mathcal{M}$
(not shown) has similar properties. Signaling is present for both
observers under both measurement settings, and in all cases the signaling
is the minimum possible. Furthermore this signaling is aligned: Any
change in one observer's choice of observable changes the other observer's
measurement outcome only in one direction (or not at all) across all
latent states of the system. Consequently, the signaling shows up
fully in a change in the latter observer's marginal probability. In
the language of Atmanspacher and Filk \cite{AtmanspacherFilk}, this
is purely communicating signaling that transmits information between
observers.

\begin{table}
\begin{centering}
\begin{tabular}{ccccc}
\hline 
$\Pr\left[M_{q}^{c}=1\right]$ & \multicolumn{4}{c}{$q$}\tabularnewline
\cline{2-5} 
$c$ & 1 & 2 & 3 & 4\tabularnewline
\hline 
$1$ & $\tfrac{2}{3}$ & $\tfrac{2}{3}$ &  & \tabularnewline
$2$ &  & $\tfrac{1}{3}$ & $\tfrac{1}{3}$ & \tabularnewline
$3$ &  &  & $\tfrac{2}{3}$ & $\tfrac{2}{3}$\tabularnewline
$4$ & $\tfrac{1}{2}$ &  &  & $\tfrac{1}{2}$\tabularnewline
\hline 
\end{tabular}
\par\end{centering}
\caption{\label{tab:Example-2-Marginals}Marginal distributions for an example
measurement system that is inconsistently connected and M-noncontextual.
The joint distribution in each context is determined by Table \ref{tab:PR-box-correlations}.}
\end{table}
\begin{table}
\begin{centering}
\begin{tabular}{crrrrrr}
\hline 
 & \multicolumn{6}{c}{$\Lambda$}\tabularnewline
\cline{2-7} 
 & 1 & 2 & 3 & 4 & 5 & 6\tabularnewline
\hline 
$F_{1}\left(\lambda,1\right)$ & $1$ & $1$ & $1$ & $1$ & $-1$ & $-1$\tabularnewline
$F_{2}\left(\lambda,1\right)$ & $1$ & $1$ & $1$ & $1$ & $-1$ & $-1$\tabularnewline
$F_{2}\left(\lambda,2\right)$ & $1$ & $1$ & $-1$ & $-1$ & $-1$ & $-1$\tabularnewline
$F_{3}\left(\lambda,2\right)$ & $1$ & $1$ & $-1$ & $-1$ & $-1$ & $-1$\tabularnewline
$F_{3}\left(\lambda,3\right)$ & $1$ & $1$ & $-1$ & $-1$ & $1$ & $1$\tabularnewline
$F_{4}\left(\lambda,3\right)$ & $1$ & $1$ & $-1$ & $-1$ & $1$ & $1$\tabularnewline
$F_{4}\left(\lambda,4\right)$ & $1$ & $-1$ & $-1$ & $-1$ & $1$ & $1$\tabularnewline
$F_{1}\left(\lambda,4\right)$ & $-1$ & $1$ & $1$ & $1$ & $-1$ & $-1$\tabularnewline
\hline 
\end{tabular}
\par\end{centering}
\caption{\label{tab:Example-2-model}A canonical model for the measurement
system in Table \ref{tab:Example-2-Marginals}. Each entry shows the
value of $F_{q}$ given the values of $\Lambda$ and $C$. $\Lambda$
is uniformly distributed over its six values.}
\end{table}

\section{\label{sec:Relationship-to-CbD}Relationship to CbD}

The contextuality-by-default approach to contextuality is originally
motivated by the observation that the traditional approach of treating
$M_{q}^{c}$ as one and the same random variable for all contexts
$c$ is mathematically incoherent, even for consistently connected
systems \cite{DzKj16}. CbD offers a rigorous alternative to this
traditional approach through the construct of a probabilistic coupling
\cite{Thorisson}.
\begin{defn}[Probabilistic coupling]
\label{def:probabilistic-coupling}A set of jointly distributed random
variables $T=\left\{ T_{q}^{c}:q\in\mathcal{Q},c\in\mathcal{C},q\prec c\right\} $
is a coupling for the measurement system $M=\left\{ M_{q}^{c}:q\in\mathcal{Q},c\in\mathcal{C},q\prec c\right\} $
if $T^{c}\sim M^{c}$ for all contexts $c$, meaning $\Pr\left[T^{c}\right]=\mu_{c}$
as distributions on $\prod_{q\prec c}\mathcal{O}_{q}$.
\end{defn}
This definition enables the definition of contextuality to be shifted
from the mathematically unsound question of a joint distribution for
$\left\{ M_{q}^{\cdot}\right\} $ (where the superscript ``$\cdot$''
indicates the context is simply ignored) to well-defined questions
regarding the joint distribution of $\left\{ T_{q}^{c}\right\} $.
For a consistently connected system $M=\left\{ M_{q}^{c}\right\} $,
the relevant coupling (if it exists) satisfies $\Pr\left[T_{q}^{c}=T_{q}^{c'}\right]=1$
for all $c,c'\succ q$. This formalizes the idea of treating $M_{q}^{c}$
and $M_{q}^{c'}$ as the same random variable. The theory also enables
the concept of contextuality to be extended to inconsistently connected
systems, by considering couplings in which $T_{q}^{c}$ and $T_{q}^{c'}$
are not always equal. We follow the CbD 2.0 version of the theory,
which is based on multimaximal couplings \cite{DzKj17-CbD2}.
\begin{defn}[Multimaximal coupling]
For a set of stochastically unrelated random variables $M_{q}=\left\{ M_{q}^{c}:c\succ q\right\} $,
a coupling $T_{q}=\left\{ T_{q}^{c}:c\succ q\right\} $ is multimaximal
if, for all $c$ and $c'$, $\Pr\left[T_{q}^{c}=T_{q}^{c'}\right]$
is maximal among all couplings of $M_{q}$.
\end{defn}
\begin{defn}[CbD-contextuality]
A measurement system $M$ is CbD-noncontextual if there exists a
coupling $T$ for $M$ such that, for each observable $q$, $T_{q}$
is a multimaximal coupling for $M_{q}$. Otherwise, $M$ is CbD-contextual.
\end{defn}
It is easy to see that CbD-contextuality agrees with standard contextuality
for consistently connected systems. If $M$ is consistently connected,
a multimaximal coupling for $M$ is one that satisfies $\Pr\left[T_{q}^{c}=T_{q}^{c'}\right]=1$
for all $c,c'\succ q$. Such a coupling is equivalent to a joint distribution
over all of the observables (i.e., $\Pr\left[\left\{ T_{q}^{\cdot}:q\in\mathcal{Q}\right\} \right]$)
that has the distributions $\mu_{c}$ as marginals.

The main results of this section rest on a correspondence between
couplings and canonical causal models:
\begin{prop}
\label{prop:coupling-model}Given a coupling $T$ for a measurement
system $M$, there exists a canonical model $\mathcal{M}$ for $M$
such that $\Delta_{c,c'}\left(F_{q}\right)=\Pr\left[T_{q}^{c}\ne T_{q}^{c'}\right]$
for all $q$ and $c,c'\succ q$. Likewise, given a canonical model
$\mathcal{M}$ for a measurement system $M$, there exists a coupling
$T$ for $M$ such that the same relationship holds.
\end{prop}
This correspondence implies the equivalence between M-contextuality
and CbD-contextuality:
\begin{thm}
\label{thm:Mcontextual-CbDcontextual}A measurement system $M$ is
M-contextual iff it is CbD-contextual.
\end{thm}
Finally, the equivalence between M-contextuality and CbD-contextuality,
together with Theorem \ref{thm:Mnoncontextual-aligned}, provides
an interpretation of CbD-contextuality in terms of aligned models:
\begin{thm}
\label{thm:CbDcontextual-aligned}A measurement system $M$ is CbD-contextual
iff there does not exist an aligned model for $M$.
\end{thm}
Thus CbD-contextuality has a simple interpretation in terms of direct
influence, specifically that a measurement system is CbD-contextual
whenever modeling that system would require hidden direct influences
or, equivalently, hidden signaling for partitionable systems. In other
words, a CbD-contextual system is one that be modeled only by violating
the no-conspiracy principle.

\section{\label{sec:Conclusions}Conclusions}

The present results show an equivalence between three approaches to
extending contextuality analysis to inconsistently connected systems:
Cavalcanti's generalized no-fine-tuning principle \cite{cavalcanti18}
for causal probabilistic models, formalized here in terms of minimal
direct influence (M-contextuality); Cervantes and Dzhafarov's no-conspiracy
principle \cite{CervantesDz18-snowqueen}, formalized here as a restriction
to aligned causal models; and the purely mathematical framework of
CbD \cite{DzKj17-CbD2}. Thus, the requirement of multimaximal couplings
in CbD is equivalent to the assumption that direct influences (in
the causal sense used here) are no greater than needed to explain
the marginal distributions of individual observables, as well as to
the assumption that direct influences never cancel out, even partially.
Although there is a sense in which any mathematical definition is
as good as any other, the fact that three different principles all
converge on the same classification of systems into contextual and
non-contextual might be taken as a stronger indication of the likely
scientific utility of this classification.

Another main contribution of this article, beyond the correspondence
between the particular definitions analyzed, is the translation between
the CbD and model-based approaches. Under this translation, alternative
conditions on couplings correspond to alternative criteria for direct
influence. For example, the maximal couplings used in earlier versions
of CbD \cite{KuDzLa15} correspond to canonical causal models minimizing
$\Pr\left[\left\{ \lambda:\exists c,c'\succ q\left(F_{q}\left(\lambda,c\right)\neq F_{q}\left(\lambda,c'\right)\right)\right\} \right]$
for each observable $q$. Likewise, other measures of causal influence
in probabilistic models \cite{janzig-etal} might translate to interesting
criteria on couplings. This strategy of translation might enable insights
from either framework to inform development in the other, to better
understand the properties of contextual inconsistently connected systems,
or to devise more refined definitions. 

Although the translation between canonical models and couplings is
fairly trivial, with the latent state of the model corresponding to
the sample space of the coupling (Proposition \ref{prop:coupling-model}),
the model-based approach offers a number of conceptual advantages.
First, it maintains the standard interpretation of contextuality as
the impossibility of explaining a system classically, meaning with
a hidden-variables theory in which measurements that were not made
are nevertheless well-defined \cite{EPR}. Second, it distinguishes
direct influence, a property of the theoretical data-generating process,
from inconsistent connectedness, a property of the data distribution
\cite{cavalcanti18}. Consequently, third, it provides a formalism
for expressing how an observable might directly depend on context,
thus enabling potential extensions wherein theoretical assumptions
regarding the physical system impose additional constraints on this
dependence.

Expanding on the third point, the definition of M-contextuality is
based on causal models that may embody rich theories of the physical
system underlying a set of measurements. Because the definition quantifies
over all such models, it depends only on the measurements themselves
(i.e., $\left\{ \mu_{c}\right\} $ or samples therefrom). It is thus
a meta-theory, providing conditions regarding what types of models
are and are not mathematically possible for a given system. However,
the model-based approach can also accommodate theoretical considerations
directly. That is, one could incorporate domain-specific assumptions
regarding direct influences, based on theoretical principles applicable
to the physical system under study, and define a system as contextual
whenever any model of that system would require stronger direct influences
than allowed by the assumed theory. This is the standard logic with
Bell scenarios, where the underlying theory is special relativity
and the constraint is that signaling between spacelike-separated observers
must be absent. The present approach can also accommodate more nuanced
constraints, including limits on the magnitude of direct influence
or signaling even when it is allowed to be nonzero. A further generalization
of the analysis presented here would be to consider model architectures
other than the canonical and partitioned ones. For example, rather
than allowing direct influence of context on each observable, $C\in Pa\left(F_{q}\right)$,
or signaling between observers, $C_{k}\in Pa\left(\tilde{F}_{k'}\right)$,
one could consider models with direct influence between observables,
$F_{q}\in Pa\left(F_{q'}\right)$, to see what additional insight
they might provide into the mathematical nature of contextuality.

As a fourth advantage, the model-based approach is naturally suited
to the fact that contextuality analysis of an empirical measurement
system is analysis of a finite dataset. As with any situation of hypothesis
testing based on sample data, it requires statistical inference, in
this case inference of whether the data could have been generated
by a model from some class. As noted in Section \ref{sec:Standard-Contextuality},
the data form a set $\hat{M}=\left\{ M_{q}^{c,i}:q\in\mathcal{Q},c\in\mathcal{C},q\prec c,1\le i\leq n_{c}\right\} $,
with each $M_{\cdot}^{c,i}$ a sample from an unknown distribution
$\mu_{c}$. The set of possibilities for $\left\{ \mu_{c}:c\in\mathcal{C}\right\} $
is a product of $\left(m_{c}-1\right)$-simplices, where $m_{c}=\left|\left\{ q:q\prec c\right\} \right|$
is the number of observables measured in context $c$. The subset
of possibilities consistent with a context-free model (or equivalently
a partitioned model with no signaling) is defined by a set of linear
inequalities and hence is a polytope within that space. Likewise,
the subset of possibilities consistent with an aligned model (canonical
or partitioned) is defined by a different polytope, containing the
first. Given a dataset $\hat{M}$, the questions of contextuality,
M-contextuality, and CbD-contextuality are questions of whether $\hat{M}$
is a sample from a set of true distributions lying in the appropriate
polytope, which can be answered with existing inferential methods
\cite{DavisStober09,HeckDavisStober,MyungEtal05}. Thus the model-based
approach can accommodate sampling error in a principled way, without
conflating it with direct influence.

The present results help to clarify and qualify Atmanspacher and Filk's
argument that CbD-contextuality is an inadequate definition because
it accounts only for communicating and not non-communicating (hidden)
signaling, whereas ``signaling whatsoever, \textquotedblleft hidden\textquotedblright{}
or not, cannot create true quantum contextuality'' \cite{AtmanspacherFilk}.
Expositions of CbD make clear that this distinction is not meaningful
within that framework, since it defines direct influence solely in
distributional terms \cite{CervantesDz18-snowqueen,DzKj16}. Thus
the two groups are essentially speaking different languages. The present
approach bridges this gap, by distinguishing the probabilistic concept
of inconsistent connectedness from the causal concept of direct influence,
and by showing how one can translate between probabilistic couplings
and causal models (Proposition \ref{prop:coupling-model}). Using
this translation, Theorem \ref{thm:CbDcontextual-aligned} shows Atmanspacher
and Filk's statement is correct: CbD-contextuality implies that a
system cannot be explained entirely by communicating (aligned) direct
influence or signaling, but it does not imply the system cannot be
explained by hidden influences or signaling. On the other hand, the
latter observation is not specific to CbD but applies to all possible
theories of contextuality: If unconstrained, direct influence can
always explain any pattern of data (Proposition \ref{prop:universality-measurement-models}),
as Bohmian mechanics demonstrates for EPR-Bell scenarios \cite{bohm52}.
Therefore, contextuality is of interest only under some restriction
on direct influence. The no-conspiracy principle is one such restriction.
In that regard, Theorem \ref{thm:CbDcontextual-aligned} unifies the
positions of Atmanspacher and Filk and of Dzhafarov, Kujala, and colleagues:
CbD-contextuality is equivalent to the proposition that a system cannot
be explained by direct influence alone, if one excludes hidden influences
a priori. The conceptual justification and scientific utility of this
exclusion will likely be a matter of further debate.

\appendix

\section*{Relating Causal and Probabilistic Approaches to Contextuality}

\subsection*{\begin{center}Supplementary Material\end{center}}

\author{\begin{center}Matt Jones (mcj@colorado.edu)\end{center}}

\author{\begin{center}University of Colorado Boulder\end{center}}

\paragraph*{\setcounter{page}{1}Proof of Proposition \ref{prop:universality-measurement-models}}

For each $c$, extend $\mu_{c}$ to a probability measure $\mu^{\left(c\right)}$
on $\prod_{q\in\mathcal{Q}}\mathcal{O}_{q}$, for example by choosing
an arbitrary distribution for each observable $q\nprec c$ and then
taking the product measure. Identify the set of possible values for
$\Lambda$ (i.e., the space of hidden states) with the Cartesian product
$\prod_{c\in\mathcal{C}}\left(\prod_{q\in\mathcal{Q}}\mathcal{O}_{q}\right)$,
and let $\mu_{\Lambda}$ be the product measure on this space obtained
from the measures $\left\{ \mu^{\left(c\right)}:c\in\mathcal{C}\right\} $.
Finally, define $F_{q}\left(\lambda,c\right)=\lambda_{c,q}$. For
any $c$, the conditional distribution $\Pr\left[\left\{ F_{q}:q\prec c\right\} \middle\vert C=c\right]$
is equal to the distribution obtained from $\mu_{\Lambda}$ by projecting
$\prod_{c'\in\mathcal{C}}\left(\prod_{q\in\mathcal{Q}}\mathcal{O}_{q}\right)\rightarrow\prod_{q\prec c}\mathcal{O}_{q}$
(taking copy $c$ from the outside product and marginalizing over
all $q\nprec c$ in the inside product), which by construction is
$\mu_{c}$.

\paragraph*{Proof of Proposition \ref{prop:noncontextual-contextfree}}

Sufficiency: Noncontextuality of $M$ implies there exists a distribution
$\mu$ over all the observables such that its projection to the observables
within each context $c$ equals the distribution $\mu_{c}$. Define
a context-free model with $\Lambda$ ranging over $\prod_{q\in\mathcal{Q}}\mathcal{O}_{q}$
with probability measure $\mu_{\Lambda}=\mu$, and with $F_{q}\left(\lambda\right)=\lambda_{q}$
for every $q$ and $\lambda$. Then the joint distribution $\Pr\left[\left\{ F_{q}:q\in\mathcal{Q}\right\} \right]$
equals $\mu$, and thus for any context $c$, the distribution $\Pr\left[\left\{ F_{q}:q\prec c\right\} \right]$
equals $\mu_{c}$. 

Necessity: If $\mathcal{M}$ is a context-free model for $M$, then
$\mu_{c}$ is the same distribution as $\Pr\left[\left\{ F_{q}:q\prec c\right\} \middle\vert C=c\right]$,
which in turn is the same as $\Pr\left[\left\{ F_{q}:q\prec c\right\} \right]$
because the $F_{q}$ do not depend on $C$. Therefore $\mu_{c}$ equals
the projection of $\Pr\left[\left\{ F_{q}:q\in\mathcal{Q}\right\} \right]$
to $\prod_{q\prec c}\mathcal{O}_{q}$ for all $c$, implying $M$
is noncontextual.

\paragraph*{Proof of Proposition \ref{prop:CC-contextfree}}

In any context-free model for $M_{q}$, $F_{q}$ is independent of
$C$, implying $\Pr\left[F_{q}\middle\vert C=c\right]=\Pr\left[F_{q}\right]$
(as distributions on $\mathcal{O}_{q}$) for all $c$. Therefore $M_{q}^{c}$
has the same distribution for all $c$, implying consistent connectedness.
Conversely, if $M$ is consistently connected then for each $q$ we
can define the probability measure $\mu_{q}$ on $\mathcal{O}_{q}$
that is the distribution shared by all $M_{q}^{c}$. A model of $M_{q}$
is then trivially constructed by letting $\Lambda$ range over $\mathcal{O}_{q}$
with distribution $\mu_{q}$ and taking $F_{q}\left(\lambda\right)=\lambda$
for all $\lambda\in\mathcal{O}_{q}$.

\paragraph*{General Definition of Aligned Canonical Models and Hidden Influences}

Given an observable $q$ with arbitrary outcome space $\mathcal{O}_{q}$
and two contexts $c,c'\succ q$, a canonical causal model $\mathcal{M}$
is said to have hidden direct influences with respect to $\left\{ q,c,c'\right\} $
when there exists a measurable set $E\subset\mathcal{O}_{q}$ such
that $\Pr\left[\left\{ \lambda:F_{q}\left(\lambda,c\right)\in E\right\} \right]>0$,
$\Pr\left[\left\{ \lambda:F_{q}\left(\lambda,c'\right)\in E\right\} \right]>0$,
and for every measurable subset $E'\subset E$, either $\Pr\left[\left\{ \lambda:F_{q}\left(\lambda,c\right)\in E',F_{q}\left(\lambda,c'\right)\notin E'\right\} \right]>0$
and $\Pr\left[\left\{ \lambda:F_{q}\left(\lambda,c\right)\notin E',F_{q}\left(\lambda,c'\right)\in E'\right\} \right]>0$,
or else $\Pr\left[\left\{ \lambda:F_{q}\left(\lambda,c\right)\in E'\right\} \right]=\Pr\left[\left\{ \lambda:F_{q}\left(\lambda,c'\right)\right.\right.$
$\left.\left.\in E'\right]\right\} =0$. A model is aligned if it
has no hidden direct influences for any $q,c,c'$. This definition
is equivalent to Definition \ref{def:hidden-influence} in the main
text when $\mathcal{O}_{q}$ is discrete, as can be seen by identifying
$E$ with $\left\{ v\right\} $.

\paragraph*{Proof of Theorem \ref{thm:Mcontextual-contextual}}

Let $M$ be a consistently connected measurement system. By Proposition
\ref{prop:CC-contextfree}, for each $q$ there exists a context-free
model $\mathcal{M}_{q}$ for $M_{q}$. The model $\mathcal{M}_{q}$
satisfies $\Delta_{c,c'}\left(F_{q}\right)=0$ for all $c,c'\succ q$,
and it can be arbitrarily extended to a model for the full system.
Therefore, $M$ is M-noncontextual iff there exists a model for $M$
with all direct influences equal to zero.

If there exists a context-free model for $M$, all direct influences
in this model are zero and therefore $M$ is M-noncontextual. Conversely,
assume $M$ is M-noncontextual a let $\mathcal{M}$ be a model for
$M$ with all direct influences equal to zero. For each $q$ and contexts
$c,c'\succ q$, define $E_{q}^{cc'}=\left\{ \lambda:F_{q}\left(\lambda,c\right)=F_{q}\left(\lambda,c'\right)\right\} $.
By assumption, $\Pr\left[E_{q}^{cc'}\right]=1$. Because $\mathcal{Q}$
and $\mathcal{C}$ are assumed to be countable, $\Pr\left[E\right]=1$,
where $E=\bigcap_{q,c,c':q\prec c,c'}E_{q}^{cc'}$. Now define a new
model $\mathcal{M}'$ by restricting the range of $\Lambda$ and the
domain of every $F_{q}$ (in the first argument) to $E$. By construction,
$\mathcal{M}'$ is a context-free model for $M$.

\paragraph*{Proof of Theorem \ref{thm:Mnoncontextual-aligned}}

Fix and $q$ and $c,c\succ q$, and let $\mu^{c}$ and $\mu^{c'}$
respectively be the distributions of $M_{q}^{c}$ and $M_{q}^{c'}$,
as probability measures on $\mathcal{O}_{q}$. By the Hahn-Jordan
decomposition theorem applied to the signed measure $\mu^{c}-\mu^{c'}$,
there exist a partition of the outcome space $\mathcal{O}_{q}=\mathcal{O}_{q}^{+}\sqcup\mathcal{O}_{q}^{-}$
and positive measures $\mu^{+}$ and $\mu^{-}$ such that $\mu^{+}\left(\mathcal{O}_{q}^{-}\right)=\mu^{-}\left(\mathcal{O}_{q}^{+}\right)=0$
and $\mu^{c}-\mu^{c'}=\mu^{+}-\mu^{-}$. Moreover, $\mu^{+}$ and
$\mu^{-}$ are unique. Define $\mu^{0}=\mu^{c}-\mu^{+}=\mu^{c'}-\mu^{-}$,
which is necessarily a positive measure, and define $\alpha=\mu^{0}\left(\mathcal{O}_{q}\right)$.
We prove the following three statements: 
\begin{enumerate}
\item The minimal direct influence across all models for $M$ is given by
$\min_{\mathcal{M}}\Delta_{c,c'}\left(F_{q}\right)=1-\alpha$.
\item If a model $\mathcal{M}$ for $M$ satisfies $\Delta_{c,c'}\left(F_{q}\right)=1-\alpha$,
then it contains no hidden influences with respect to $\left\{ q,c,c'\right\} $.
\item Conversely, if a model $\mathcal{M}$ for $M$ contains no hidden
influences with respect to $\left\{ q,c,c'\right\} $, then it satisfies
$\Delta_{c,c'}\left(F_{q}\right)=1-\alpha$ .
\end{enumerate}
Together, these three statements imply that any model $\mathcal{M}$
for $M$ is aligned iff it minimizes all direct influences, which
in turn implies the theorem.

\uline{Proof of Statement (i)}. Let $\mathcal{M}$ be any canonical
model for $M$. The direct influence in $\mathcal{M}$ is constrained
by
\begin{align*}
\Delta_{c,c'}\left(F_{q}\right) & \ge\Pr\left[\left\{ \lambda:F_{q}\left(\lambda,c\right)\in\mathcal{O}_{q}^{+},F_{q}\left(\lambda,c'\right)\notin\mathcal{O}_{q}^{+}\right\} \right]\\
 & \ge\Pr\left[\left\{ \lambda:F_{q}\left(\lambda,c\right)\in\mathcal{O}_{q}^{+}\right\} \right]-\Pr\left[\left\{ \lambda:F_{q}\left(\lambda,c'\right)\in\mathcal{O}_{q}^{+}\right\} \right]\\
 & =\mu^{c}\left(\mathcal{O}_{q}^{+}\right)-\mu^{c'}\left(\mathcal{O}_{q}^{+}\right)\\
 & =\mu^{+}\left(\mathcal{O}_{q}^{+}\right)-\mu^{-}\left(\mathcal{O}_{q}^{+}\right)\\
 & =\mu^{+}\left(\mathcal{O}_{q}\right)\\
 & =\mu^{c}\left(\mathcal{O}_{q}\right)-\mu^{0}\left(\mathcal{O}_{q}\right)\\
 & =1-\alpha.
\end{align*}
Therefore $1-\alpha$ is a lower bound for $\Delta_{c,c'}\left(F_{q}\right)$.
To construct a model meeting this bound, let $\Lambda$ range over
$\mathcal{O}_{q}\times\mathcal{O}_{q}$ and define $F_{q}\left(\left(v_{1},v_{2}\right),c\right)=v_{1}$
and $F_{q}\left(\left(v_{1},v_{2}\right),c'\right)=v_{2}$ for all
$v_{1},v_{2}\in\mathcal{O}_{q}$. Let $\pi^{d}:\mathcal{O}_{q}\to\mathcal{O}_{q}\times\mathcal{O}_{q}$
be the diagonal embedding $\pi^{d}\left(v\right)=\left(v,v\right)$,
and define the push-forward measure $\mu^{d}=\pi_{*}^{d}\left(\mu^{0}\right)$,
so that $\mu^{d}\left(E\right)=\mu^{0}\left(\left\{ v\in\mathcal{O}_{q}:\left(v,v\right)\in E\right\} \right)$
for all measurable $E\subset\mathcal{O}_{q}\times\mathcal{O}_{q}$.
Define a second measure $\mu^{u}$ on $\mathcal{O}_{q}\times\mathcal{O}_{q}$,
generated by 
\[
\mu^{u}\left(E_{1}\times E_{2}\right)=\frac{\mu^{+}\left(E_{1}\right)\cdot\mu^{-}\left(E_{2}\right)}{1-\alpha}
\]
for all measurable $E_{1},E_{2}\subset\mathcal{O}_{q}$. Now define
the distribution on $\Lambda$ by $\Pr\left[\Lambda\right]=\mu^{d}+\mu^{u}$.
For any measurable $E\subset\mathcal{O}_{q}$, 
\begin{align*}
\Pr\left[F_{q}\in E\middle\vert C=c\right] & =\mu^{d}\left(E\times\mathcal{O}_{q}\right)+\mu^{u}\left(E\times\mathcal{O}_{q}\right)\\
 & =\mu^{0}\left(E\right)+\frac{\mu^{+}\left(E\right)\cdot\mu^{-}\left(\mathcal{O}_{q}\right)}{1-\alpha}\\
 & =\mu^{0}\left(E\right)+\frac{\mu^{+}\left(E\right)\cdot\left(\mu^{c'}\!\left(\mathcal{O}_{q}\right)-\mu^{0}\left(\mathcal{O}_{q}\right)\right)}{1-\alpha}\\
 & =\mu^{c}\left(E\right).
\end{align*}
A similar calculation shows $\Pr\left[F_{q}\in E\middle\vert C=c'\right]=\mu^{c'}\!\left(E\right)$.
Therefore $\mathcal{M}$ is a model for the subsystem $\left\{ M_{q}^{c},M_{q}^{c'}\right\} $,
which can be arbitrarily extended to a model for the full system $M$.
The direct influence is given by
\begin{align*}
\Delta_{c,c'}\left(F_{q}\right) & =\mu^{d}\left(\mathcal{O}_{q}\times\mathcal{O}_{q}\right)\\
 & =\frac{\left(\mu^{c}\left(\mathcal{O}_{q}\right)-\mu^{0}\left(\mathcal{O}_{q}\right)\right)\cdot\left(\mu^{c'}\!\left(\mathcal{O}_{q}\right)-\mu^{0}\left(\mathcal{O}_{q}\right)\right)}{1-\alpha}\\
 & =1-\alpha.
\end{align*}

\uline{Proof of Statement (ii)}. Assume $\mathcal{M}$ has hidden
influences with respect to $\left\{ q,c,c'\right\} $, and let $E$
be as given above in the General Definition of Hidden Influences.
Define $E^{+}=E\cap\mathcal{O}_{q}^{+}$ and $E^{-}=E\cap\mathcal{O}_{q}^{-}$.
Because $\Pr\left[\left\{ \lambda:F_{q}\left(\lambda,c\right)\in E\right\} \right]>0$
and $\Pr\left[\left\{ \lambda:F_{q}\left(\lambda,c'\right)\in E\right\} \right]>0$,
it cannot be that $\Pr\left[\left\{ \lambda:F_{q}\left(\lambda,c\right)\in E^{+}\right\} \right]=\Pr\left[\left\{ \lambda:F_{q}\left(\lambda,c'\right)\in E^{+}\right\} \right]=0$
and $\Pr\left[\left\{ \lambda:F_{q}\left(\lambda,c\right)\in E^{-}\right\} \right]=\Pr\left[\left\{ \lambda:F_{q}\left(\lambda,c'\right)\in E^{-}\right\} \right]=0$.
Without loss of generality, assume the former equality, $\Pr\left[\left\{ \lambda:F_{q}\left(\lambda,c\right)\in E^{+}\right\} \right]=\Pr\left[\left\{ \lambda:F_{q}\left(\lambda,c'\right)\in E^{+}\right\} \right]=0$,
is false. Then the definition of hidden influences implies $\Pr\left[\left\{ \lambda:F_{q}\left(\lambda,c\right)\notin E^{+},F_{q}\left(\lambda,c'\right)\in E^{+}\right\} \right]>0$.
Because $E^{+}\subset\mathcal{O}_{q}^{+}$, the sets $\left\{ \lambda:F_{q}\left(\lambda,c\right)\in\mathcal{O}_{q}^{+},F_{q}\left(\lambda,c'\right)\notin\mathcal{O}_{q}^{+}\right\} $
and $\left\{ \lambda:F_{q}\left(\lambda,c\right)\notin E^{+},F_{q}\left(\lambda,c'\right)\in E^{+}\right\} $
are disjoint, so we can bound the direct influence as $\Delta_{c,c'}\left(F_{q}\right)\ge\Pr\left[\left\{ \lambda:F_{q}\left(\lambda,c\right)\in\mathcal{O}_{q}^{+},F_{q}\left(\lambda,c'\right)\notin\mathcal{O}_{q}^{+}\right\} \right]+\Pr\left[\left\{ \lambda:F_{q}\left(\lambda,c\right)\notin E^{+},F_{q}\left(\lambda,c'\right)\in E^{+}\right\} \right]$.
The proof of Statement 1 shows the former of these probabilities is
at least $1-\alpha$, and therefore we have $\Delta_{c,c'}\left(F_{q}\right)>1-\alpha$.
Thus we have shown any model with hidden influences cannot satisfy
$\Delta_{c,c'}\left(F_{q}\right)=1-\alpha$.

\uline{Proof of Statement (iii)}. We first prove that $\Pr\left[\left\{ \lambda:F_{q}\left(\lambda,c\right)\notin E,F_{q}\left(\lambda,c'\right)\in E\right\} \right]=0$
for any measurable $E\subset\mathcal{O}_{q}^{+}$. To see this, assume
the contrary, that $\Pr\left[\left\{ \lambda:F_{q}\left(\lambda,c\right)\notin E,F_{q}\left(\lambda,c'\right)\in E\right\} \right]=\varepsilon$
with $\varepsilon>0$ for some $E\subset\mathcal{O}_{q}^{+}$. Using
alignment of $\mathcal{M}$, the probability $\varepsilon$ can be
squeezed into successively smaller subsets of $E$ so as to produce
a contradiction. Specifically, define a property $S$ with $S\left(E'\right)$
being the statement that $E'$ is a measurable subset of $E$ with
$\Pr\left[\left\{ \lambda:F_{q}\left(\lambda,c\right)\notin E,F_{q}\left(\lambda,c'\right)\in E\setminus E'\right\} \right]=0$.
Note that $S$ is preserved under countable intersection and that
$S\left(E'\right)$ implies $\Pr\left[\left\{ \lambda:F_{q}\left(\lambda,c\right)\notin E,F_{q}\left(\lambda,c'\right)\in E'\right\} \right]=\varepsilon$.
If we define $\beta=\inf\left\{ \Pr\left[\left\{ \lambda:F_{q}\left(\lambda,c\right)\in E'\right\} \right]:S\left(E'\right)\right\} $,
then the countable intersection property just stated implies there
exists a set $E_{0}\subset E$ meeting this bound: $\Pr\left[\left\{ \lambda:F_{q}\left(\lambda,c\right)\in E_{0}\right\} \right]=\beta$
and $\Pr\left[\left\{ \lambda:F_{q}\left(\lambda,c\right)\notin E,F_{q}\left(\lambda,c'\right)\in E_{0}\right\} \right]=\varepsilon$.
If $\beta>0$, then alignment of $\mathcal{M}$ implies there are
no hidden influences within $E_{0}$; that is, there exists $E_{1}\subset E_{0}$
such that $\Pr\left[\left\{ \lambda:F_{q}\left(\lambda,c\right)\in E_{1}\right\} \right]>0$
or $\Pr\left[\left\{ \lambda:F_{q}\left(\lambda,c'\right)\in E_{1}\right\} \right]>0$,
and also $\Pr\left[\left\{ \lambda:F_{q}\left(\lambda,c\right)\notin E_{1},F_{q}\left(\lambda,c'\right)\in E_{1}\right\} \right]=0$
or $\Pr\left[\left\{ \lambda:F_{q}\left(\lambda,c\right)\in E_{1},F_{q}\left(\lambda,c'\right)\notin E_{1}\right\} \right]=0$.
Because $E_{1}\subset\mathcal{O}_{q}^{+}$, $\Pr\left[\left\{ \lambda:F_{q}\left(\lambda,c\right)\in E_{1}\right\} \right]\ge\Pr\left[\left\{ \lambda:\right.\right.$
$\left.\left.F_{q}\left(\lambda,c'\right)\in E_{1}\right\} \right]$,
which implies that the former relation in each of the two disjunctions
just given holds: $\Pr\left[\left\{ \lambda:F_{q}\left(\lambda,c\right)\in E_{1}\right\} \right]>0$
and $\Pr\left[\left\{ \lambda:F_{q}\left(\lambda,c\right)\notin E_{1},F_{q}\left(\lambda,c'\right)\in E_{1}\right\} \right]=0$.
This in turn implies $S\left(E_{0}\setminus E_{1}\right)$ and $\Pr\left[\left\{ \lambda:F_{q}\left(\lambda,c\right)\in E_{0}\setminus E_{1}\right\} \right]<\beta$,
contradicting the definition of $\beta$. On the other hand, if $\beta=0$
then $\Pr\left[\left\{ \lambda:F_{q}\left(\lambda,c\right)\in E_{0}\right\} \right]<\Pr\left[\left\{ \lambda:F_{q}\left(\lambda,c'\right)\in E_{0}\right\} \right]$,
contradicting the fact that $E_{0}\subset\mathcal{O}_{q}^{+}$. Therefore
the supposed set $E\subset\mathcal{O}_{q}^{+}$ with $\Pr\left[\left\{ \lambda:F_{q}\left(\lambda,c\right)\notin E,F_{q}\left(\lambda,c'\right)\in E\right\} \right]$
$>0$ cannot exist.

Next, let $\left(E_{n}\right)_{n\in\mathbb{N}}$ be a countable basis
for $\mathcal{O}_{q}^{+}$, using the assumption that $\mathcal{O}_{q}$
is second-countable. Take any $v_{1}\in\mathcal{O}_{q}$ and $v_{2}\in\mathcal{O}_{q}^{+}$
with $v_{1}\ne v_{2}$. Because $\mathcal{O}_{q}$ is Hausdorff, there
exists be an open neighborhood $N$ of $v_{2}$ not containing $v_{1}$.
Because $\left(E_{n}\right)_{n\in\mathbb{N}}$ is a basis for $\mathcal{O}_{q}^{+}$,
there exists some $E_{m}$ with $v_{2}\in E_{m}\subset N\cap\mathcal{O}_{q}^{+}$
and hence also $v_{1}\notin E_{m}$. This shows that $\left\{ \lambda:F_{q}\left(\lambda,c'\right)\in\mathcal{O}_{q}^{+},F_{q}\left(\lambda,c\right)\ne F_{q}\left(\lambda,c'\right)\right\} $
is a subset of $\bigcup_{n}\left\{ \lambda:F_{q}\left(\lambda,c\right)\notin E_{n},F_{q}\left(\lambda,c'\right)\in E_{n}\right\} $.
Therefore 
\begin{align*}
\Pr\left[\left\{ \lambda:F_{q}\left(\lambda,c'\right)\in\mathcal{O}_{q}^{+},F_{q}\left(\lambda,c\right)\ne F_{q}\left(\lambda,c'\right)\right\} \right] & \le\sum_{n}\Pr\left[\left\{ \lambda:F_{q}\left(\lambda,c\right)\notin E_{n},F_{q}\left(\lambda,c'\right)\in E_{n}\right\} \right]\\
 & =0,
\end{align*}
which in turn implies
\begin{align*}
\Pr\left[\left\{ \lambda:F_{q}\left(\lambda,c\right)=F_{q}\left(\lambda,c'\right)\in\mathcal{O}_{q}^{+}\right\} \right] & =\Pr\left[\left\{ \lambda:F_{q}\left(\lambda,c'\right)\in\mathcal{O}_{q}^{+}\right\} \right]\\
 & =\mu^{0}\left(\mathcal{O}_{q}^{+}\right).
\end{align*}
A parallel argument shows $\Pr\left[\left\{ \lambda:F_{q}\left(\lambda,c\right)=F_{q}\left(\lambda,c'\right)\in\mathcal{O}_{q}^{-}\right\} \right]=\mu^{0}\left(\mathcal{O}_{q}^{-}\right)$.
Therefore the total direct influence in $\mathcal{M}$ for $q,c,c'$
is given by
\begin{align*}
\Delta_{c,c'}\left(F_{q}\right) & =1-\Pr\left[\left\{ \lambda:F_{q}\left(\lambda,c\right)=F_{q}\left(\lambda,c'\right)\right\} \right]\\
 & =1-\Pr\left[\left\{ \lambda:F_{q}\left(\lambda,c\right)=F_{q}\left(\lambda,c'\right)\in\mathcal{O}_{q}^{+}\right\} \right]-\Pr\left[\left\{ \lambda:F_{q}\left(\lambda,c\right)=F_{q}\left(\lambda,c'\right)\in\mathcal{O}_{q}^{-}\right\} \right]\\
 & =1-\mu^{0}\left(\mathcal{O}_{q}^{+}\right)-\mu^{0}\left(\mathcal{O}_{q}^{-}\right)\\
 & =1-\alpha.
\end{align*}

\paragraph*{General Definition of Aligned Partitioned Models and Hidden Signals}

Given an observer $k$, an observable $q\in\mathcal{Q}_{k}$ with
arbitrary outcome space $\mathcal{O}_{q}$, and contexts $c$ and
$c'$ with $c_{k}=c'_{k}=q$, a partitioned model $\tilde{\mathcal{M}}$
is said to have hidden signals with respect to $\left\{ k,c,c'\right\} $
when there exists a measurable set $E\subset\mathcal{O}_{q}$ such
that $\Pr\left[\left\{ \lambda:\tilde{F}_{k}\left(\lambda,c\right)\in E\right\} \right]>0$,
$\Pr\left[\left\{ \lambda:\tilde{F}_{k}\left(\lambda,c'\right)\in E\right\} \right]>0$,
and for every measurable subset $E'\subset E$, either $\Pr\left[\left\{ \lambda:\tilde{F}_{k}\left(\lambda,c\right)\in E',\tilde{F}_{k}\left(\lambda,c'\right)\notin E'\right\} \right]>0$
and $\Pr\left[\left\{ \lambda:\tilde{F}_{k}\left(\lambda,c\right)\notin E',\tilde{F}_{k}\left(\lambda,c'\right)\in E'\right\} \right]>0$,
or else $\Pr\left[\left\{ \lambda:\tilde{F}_{k}\left(\lambda,c\right)\in E'\right\} \right]=\Pr\left[\left\{ \lambda:\tilde{F}_{k}\left(\lambda,c'\right)\right.\right.$
$\left.\left.\in E'\right\} \right]=0$. A partitioned model is aligned
if it has no hidden signals for any $k,c,c'$. This definition is
equivalent to Definition \ref{def:hidden-signals} in the main text
when $\mathcal{O}_{q}$ is discrete for all $q\in\mathcal{Q}_{k}$,
as can be seen by identifying $E$ with $\left\{ v\right\} $.

\paragraph*{Proof of Proposition \ref{prop:noncontextual-nosignaling}}

If $M$ is noncontextual, then Proposition \ref{prop:noncontextual-contextfree}
implies there exists a context-free canonical model $\mathcal{M}$
for $M$. The corresponding partitioned model $\tilde{\mathcal{M}}$
is easily seen to be a model for $M$ with no signaling. Conversely,
if there is a partitioned model $\tilde{\mathcal{M}}$ for $M$ that
has no signaling, the corresponding canonical model $\mathcal{M}$
is context-free, and Proposition \ref{prop:noncontextual-contextfree}
then implies $M$ is noncontextual.

\paragraph*{Proof of Theorem \ref{thm:partitioned-Mcontextuality}}

Let $\tilde{\mathcal{M}}$ be any partitioned model for $M$, with
$\mathcal{M}$ the corresponding canonical model. As observed in the
main text, direct influence in $\mathcal{M}$ and signaling in $\tilde{\mathcal{M}}$
exactly correspond, in that $\tilde{\Delta}_{c,c'}\left(\tilde{F}_{k}\right)=\Delta_{c,c'}\left(F_{q}\right)$
whenever $c_{k}=c'_{k}=q$. Therefore $\tilde{\mathcal{M}}$ minimizes
all signaling iff $\mathcal{M}$ minimizes all direct influences.
The theorem then follows from the definition of M-noncontextuality,
as the existence of such an $\mathcal{M}$.

\paragraph*{Proof of Theorem \ref{thm:partitionable-aligned}}

If $\tilde{\mathcal{M}}$ is an aligned partitioned model for $M$,
then the corresponding canonical model $\mathcal{M}$ is also aligned,
implying $M$ is M-noncontextual by Theorem \ref{thm:Mnoncontextual-aligned}.
Conversely, if $M$ is M-noncontextual, there exists an aligned canonical
model $\mathcal{M}$ for $M$ by Theorem \ref{thm:Mnoncontextual-aligned},
and the corresponding partitioned model $\tilde{\mathcal{M}}$ is
also aligned.

\paragraph*{Proof of Proposition \ref{prop:coupling-model}}

First part: Let $\left(\Omega,\Sigma,P\right)$ be the sample space
for the jointly distributed random variables composing $T$, such
that each $T_{q}^{c}$ is a function $\Omega\rightarrow\mathcal{O}_{q}$.
Define $\mathcal{M}$ by letting $\Lambda$ range over $\Omega$ with
distribution $P$ and defining each $F_{q}$ by $F_{q}\left(\lambda,c\right)=T_{q}^{c}\left(\lambda\right)$
for $c\succ q$ and choosing arbitrary values for $F_{q}\left(\lambda,c\right)$
for $c\nsucc q$ (for all $\lambda\in\Omega$). Then for any context
$c$ and measurable subsets $V_{q}\subset\mathcal{O}_{q}$, 
\begin{align*}
\Pr\left[\forall q\prec c\left(F_{q}\in V_{q}\right)\middle\vert C=c\right] & =\Pr\left[\left\{ \lambda:\forall q\prec c\left(F_{q}\left(\lambda,c\right)\in V_{q}\right)\right\} \right]\\
 & =\Pr\left[\left\{ \lambda:\forall q\prec c\left(T_{q}^{c}\left(\lambda\right)\in V_{q}\right)\right\} \right]\\
 & =\Pr\left[\forall q\prec c\left(T_{q}^{c}\in V_{q}\right)\right]\\
 & =\Pr\left[\forall q\prec c\left(M_{q}^{c}\in V_{q}\right)\right].
\end{align*}
Therefore $\mathcal{M}$ is a model for $M$. For any $q$ and $c,c'\succ q$,
the claimed equality holds:
\begin{align*}
\Delta_{c,c'}\left(F_{q}\right) & =\Pr\left[\left\{ \lambda:F_{q}\left(\lambda,c\right)\neq F_{q}\left(\lambda,c'\right)\right\} \right]\\
 & =\Pr\left[\left\{ \lambda:T_{q}^{c}\left(\lambda\right)\neq T_{q}^{c'}\left(\lambda\right)\right\} \right]\\
 & =\Pr\left[T_{q}^{c}\neq T_{q}^{c'}\right].
\end{align*}

Second part: Given $\mathcal{M}=\left(\Lambda,C,\left\{ F_{q}\right\} \right)$,
let $\Omega=\left\{ \lambda\right\} $ be the range of $\Lambda$
with $P=\Pr\left[\Lambda\right]$ the associated probability measure
on $\Omega$ and $\Sigma$ the sigma-algebra of measurable sets of
values for $\Lambda$. Then $\left(\Omega,\Sigma,P\right)$ defines
a sample space. For each $q$ and $c\succ q$, define a random variable
$T_{q}^{c}$ on this sample space by $T_{q}^{c}\left(\lambda\right)=F_{q}\left(\lambda,c\right)$.
Then derivations similar to those above show that $T=\left\{ T_{q}^{c}\right\} $
is a coupling for $M$ and that $\Delta_{c,c'}\left(F_{q}\right)=\Pr\left[T_{q}^{c}\ne T_{q}^{c'}\right]$
for all $q$ and $c,c'\succ q$.

\paragraph*{Proof of Theorem \ref{thm:Mcontextual-CbDcontextual}}

If $M$ is M-noncontextual, then there exists a canonical causal model
$\mathcal{M}$ for $M$ that simultaneously minimizes all direct influences.
The corresponding coupling $T$ provided by Proposition \ref{prop:coupling-model}
minimizes $\Pr\left[T_{q}^{c}\ne T_{q}^{c'}\right]$ for all $q$
and $c,c'\succ q$. Therefore $T_{q}$ is multimaximal for all $q$,
implying $M$ is CbD-noncontextual. Conversely, if $M$ is CbD-noncontextual
then there exists a coupling $T$ for $M$ such that $T_{q}$ is multimaximal
for all $q$, implying $\Pr\left[T_{q}^{c}\ne T_{q}^{c'}\right]$
is minimal for all $c,c'\succ q$. The corresponding canonical model
$\mathcal{M}$ provided by Proposition \ref{prop:coupling-model}
minimizes $\Delta_{c,c'}\left(F_{q}\right)$ for all $q$ and $c,c'\succ q$,
implying $M$ is M-noncontextual.

\paragraph*{Proof of Theorem \ref{thm:CbDcontextual-aligned}}

The theorem follows directly from Theorems \ref{thm:Mnoncontextual-aligned}
and \ref{thm:Mcontextual-CbDcontextual}: CbD-contextuality is equivalent
to M-contextuality, which is equivalent to the non-existence of an
aligned model.
\end{document}